\documentclass{article}
\begin{document}
\title{
AN AUTOMATIC SYSTEM TO DISCRIMINATE MALIGNANT FROM BENIGN MASSIVE LESIONS IN MAMMOGRAMS
}
\author{
P. Delogu$^{1,2}$, M.E. Fantacci$^{1,2}$, P. Kasae$^{3}$ and A. Retico$^{2}$\\
$^{1}${\em Dipartimento di Fisica dell'Universit\`a di Pisa,} \\
{\em Largo Pontecorvo 3, 56127 Pisa, Italy}\\ 
$^{2}${\em Istituto Nazionale di Fisica Nucleare, Sezione di Pisa,} \\
{\em Largo Pontecorvo 3, 56127 Pisa, Italy}\\
$^{3}${\em The Abdus Salam International Center for Theoretical Physics,}\\
{\em Strada Costiera 11, P.O. Box 563, I-34100 Trieste, Italy}
}
\date{}
\maketitle
\baselineskip=11.6pt
\begin{abstract}
Evaluating the degree of   malignancy of a massive lesion on the basis of the mere visual analysis of the mammogram is a non-trivial task. We developed a semi-automated system for massive-lesion characterization with the aim to support the radiological diagnosis. 
A dataset of 226 masses has been used in the present analysis. The system performances have been evaluated in terms of the area under the ROC curve, obtaining  $A_z=0.80\pm 0.04$. 
\end{abstract}
\baselineskip=14pt
\section{Introduction}
Breast cancer is still one of the most common forms of cancer among women, despite a significant decrease has occurred in the breast cancer mortality in the last few decades~\cite{Landis}. Mammography is widely recognized as the most reliable technique for early detection of this pathology. However, characterizing the massive lesion malignancy by means exclusively of a visual analysis of the mammogram is an extremely difficult task and a high number of unnecessary biopsies are actually performed in the routine clinical activity. Computerized methods have recently shown a great potential in assisting radiologists in the malignant vs. benign decision, by providing them with a second opinion about the visual diagnosis of the lesion.
\section{Method}
The computer-aided diagnosis (CADi) system we present is based on a three-stage algorithm: 1) a segmentation technique identifies the contour of the massive lesion on the mammogram; 2) several features, based on size and shape of the lesion, are computed; 3) a neural classifier analyzes the features and outputs a likelihood of malignancy for that lesion. 
The segmentation method is a gradient-based one: it is able to identify the mass boundaries inside a physician-located region of interest (ROI) image. The algorithm is based on the maximization of the local variance along several radial lines connecting the approximate mass center to the ROI boundary~\cite{Chen}. The critical points maximizing the local variance on each radial line are interpolated, thus a rough mass shape is identified. The procedure is iterated for each point inside the approximate mass, resulting in a more accurate identification of the mass boundary. The main advantage of this segmentation technique is that no free parameters have to be fitted on the dataset to be analyzed, thus it can in principle be directly applied to datasets acquired in different conditions without any ad-hoc modification. 
Sixteen features are computed for each segmented mass, some of them being more sensitive to the shape and some to the texture of the lesion. They are: area, perimeter, circularity, mean and standard deviation of the normalized radial length, radial distance entropy, zero crossing, maximum and minimum axes, mean and standard deviation of the variation ratio, convexity; mean, standard deviation, skewness and kurtosis of the grey-level distribution.
The features are analyzed by a multi-layered feed-forward neural network trained with the error back-propagation algorithm. The classifier performances are evaluated according to the 5$\times$2 cross validation method.

\section{Results}
\begin{figure}[t]
 \vspace{6.0cm}
\includegraphics{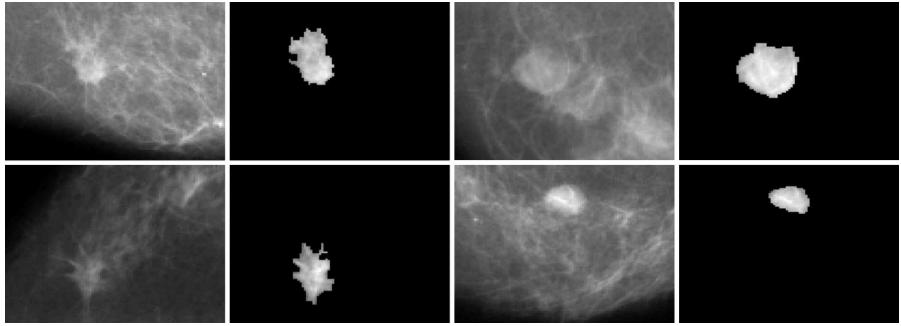}
 \caption{\it
    Examples of 
malignant (left) and benign (right) segmented masses.
    \label{fig:masses} }
\end{figure}

\begin{figure}[t]
\vspace{7.0cm}
\begin{center}
\includegraphics{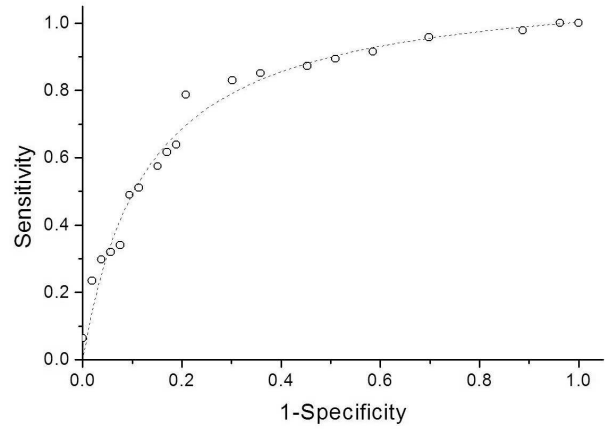}
 \caption{\it ROC curve. 
    \label{fig:FROC} }
\end{center}
\end{figure}

In this work we present the results obtained on a dataset of 226 massive lesions (109 malignant and 117 benign) extracted from a  database of mammograms collected in the framework of a collaboration between physicists from several Italian Universities and INFN  Sections, and radiologists from several Italian Senological Centers~\cite{database}.
Despite the boundaries of the masses are usually not very sharp, our segmentation procedure leads to an accurate identification of the mass shapes both in malignant and benign cases, as shown in fig.~\ref{fig:masses}.
The performances of the  neural network in classifying the features extracted from each mass have been evaluated in terms of the sensitivity and the specificity on the test sets: the average values obtained are 78.1\% and 79.1\% respectively. The discriminating capability of the system has been evaluated also in terms of the  receiver operating characteristic (ROC) analysis (see fig.~\ref{fig:FROC})~\cite{Metz}. The estimated area under the ROC curve is $A_z=0.80\pm 0.04$. 

\section{Conclusions}
Mass segmentation plays a key role in CADi systems to be used for supporting radiologists in the malignant vs. benign decision. We developed a robust technique based on edge detection to segment mass lesions from the surrounding normal tissue. The results so-far obtained in the classification of malignant and benign masses indicate that the segmentation procedure we developed provides an accurate approximation of the mass shapes and that the features we took into account for the classification have a good discriminating power.

\section{Acknowledgments}

We are grateful to Dr M.  Tonutti from Cattinara Hospital (Trieste, Italy) for her essential contribution to the present analysis.
We acknowledge Dr S. Franz from ICTP (Trieste, Italy) for useful discussions. 

\end{document}